# THE PERMUTATION ENTROPY AND ITS APPLICATIONS ON FIRE TESTS DATA

Flavia-Corina Mitroi-Symeonidis[a*]; Ion Anghel[b]; Octavian Lalu[c]; Constantin Popa[d]

[a*](corresponding author) Police Academy "Alexandru Ioan Cuza", Fire Officers Faculty, Str. Morarilor 3, Sector 2, Bucharest RO-022451, Romania. E-mail address: fcmitroi@yahoo.com

[b]Police Academy "Alexandru Ioan Cuza", Fire Officers Faculty, Str. Morarilor 3, Sector 2, Bucharest RO-022451, Romania. E-mail address: ion.anghel@academiadepolitie.ro

[c]British Research Establishment, Bucknalls Lane, WD25 9NH, Watford, UK. E-mail address: octavian.lalu@gmail.com

[d]Romanian Fire Safety Association, Bucharest, Romania. E-mail address: costi_popa001@yahoo.com

**Abstract**

Based on the data gained from a full-scale experiment, the order/disorder characteristics of the compartment fire temperatures are analyzed. Among the known permutation/encoding type entropies used to analyze time series, we look for those that fit better in the fire phenomena. The literature in its major part does not focus on time series with data collected during full-scale fire experiments, therefore we do not only perform our analysis and report the results, but also discuss methods, algorithms, the novelty of our entropic approach and details behind the scene. The embedding dimension selection in the complexity evaluation is also discussed. Finally, more research directions are proposed.

## 1. Introduction

The permutation entropy can be used as a measure of the unpredictability of the combustion at different points of interest in the fire compartment, allowing us to examine some spatial pattern variation exhibited by fire hazards. The experimental setup described in Section 2 is intended for measurements at the position of the firefighters that can help to assess the health risks of fire exposure.

Researchers have conducted a few studies using the entropic analysis of the fire phenomena, a recent approach in the literature. See (Takagi, Gotoda, Tokuda, & Miyano, 2018) and (Murayama, Kaku, Funatsu, & Gotoda, 2019). Our aim is to analyze some experimental data by the tools of the information theory. Whereas there is no universal formula for the entropy and so many are proposed in the literature, we make comparisons among several existing methods of determining the underlying probabilities and additional proposed variants. Obviously, the full-scale fire experiments provide the most credible experimental data. Pointing out abnormal values and structure of the experimental time series would indicate the usefulness of some methods, or the irrelevancy of others. We discuss the mathematical and technical causes which determine the failure of some algorithms and the advantages of using others.

Figure 1 shows the idealized fire curve of fire which describes the evolution of the temperature during a fire experiment in a compartment. The lower curve corresponds to the regime of a quasi-steady low-intensity fire. See (Graham, Makhviladze, & Roberts , 1995).

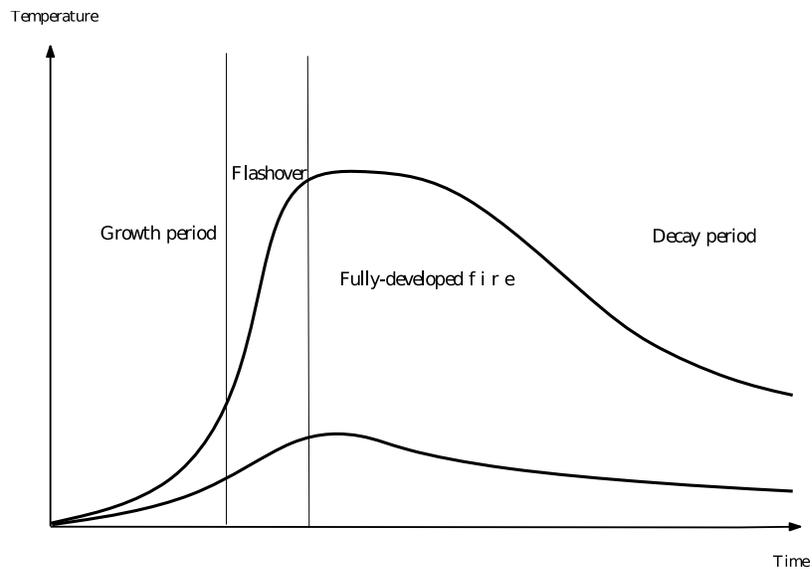

*Figure 1 – Idealized time-temperature fire curve*

The evolution of the fire depends on the shape and on the dimensions of the room, the available air supply, the insulation materials and the position of the fuel. Important information in this respect can be found in (Babrauskas, Estimating Room Flashover Potential, 1980), (McCaffrey, Quintiere, & Harkleroad, 1981), (Thomas, 1981), (Peacock, Reneke, Bukowski, & Babrauskas, 1999). Over the last decades there has been a great increase in the mathematical modeling of fire development within buildings. Models of this kind provide insight into the fundamental processes of fire development and have contributions in direct practical terms such as assessment of a specific design. (Beard, 2010)

The turbulence phenomena are characterized by random fluctuations of characteristics describing the state of the system around some average values; therefore, we analyze not only the fluctuations of the temperature values, but also their averages.

The purpose of this paper is to perform a local entropic analysis of the evolution of the temperature during a fire experiment. The next section is dedicated to the description of the experimental setup (materials and methods) and giving details on the collected data. In Section 3 we present the mathematical tools used to perform the analysis, we introduce new tools and have remarks on their properties and use for the time series analysis, followed by the main results and their interpretation. Section 4 is dedicated to conclusions and further research directions.

## 2. The fire scenario. Experimental results, materials and methods. Data acquisition

In this paper we investigate the experimental data that has been collected during a full-scale fire experiment conducted at Fire Officers Faculty in Bucharest.

Checking the last both international (Kerber, 2012) and national (IGSU, 2010) statistics, one can see that people (both fire fighters as well as the normal users of the buildings) still die in fires. A significant reason for these casualties is the lack of understanding how a residential fire behaves, given the latest changes that affect the dynamics of this type of fires. These changes involve bigger volumes of fire enclosures (rooms/buildings), different geometry, a rise in the amount of synthetic combustible materials and a change in the structure and fire reaction of the construction materials.

Thirty years ago, a classically furnished room under fire was reaching flashover point 29 minutes after fire initiation. Nowadays, as a result of the changes presented above, a similar use room reaches flashover in less than 5 minutes (Kerber, 2012). This is why fire researchers today should concentrate the efforts on studying the dynamics of fires that involve today's combustible materials (i.e. plywood, OSB, gypsum board, PVC etc.) and also on training fire fighters in safe and in as close to reality as possible conditions.

Wood is an integrated part of the load bearing structure, also it is the main source of matter used to create furniture to be found in buildings all over the world. The pyrolysis of wood starts at over 225°C and ends at temperatures below 500°C (Babrauskas, 2002). The wood produces less smoke than most of the plastic materials used today. In decent ventilation conditions, wood can produce 25-100 $m^2/kg$ of smoke, when the same amount of plastic materials releases, under the same conditions, hundreds or even thousands of $m^2/kg$ (Hakkarainen, et al., 2005).

Smoke formation is dependent on the burning material and on factors like oxygen feed and the type of combustion (e.g. with flame or incandescence).

There are two parameters affecting fire performances of wood-made products: the density and the thickness. When the density is smaller, it takes less time for the wood surface to reach the ignition temperature (which is approx. 360°C for piloted ignition of wood). Similarly, after ignition, the flame will propagate quicker, lesser the density (Mikkola, 2004).

The elevation or position (i.e. ceiling, wall, floor level) of a burning product in a fire room is of upmost importance. Especially ceilings and upper parts of walls are critical locations in comparison with lower floor levels. According to the fire reaction tests for OSB (Grexa, O. et al., 2011) (Grexa, Dietenberger, & White, 2011), the results in Table 1 below were obtained:

*Table 1 - Results of fire reaction tests (Grexa, Dietenberger, & White, 2011) for OSB (Oriented Strand Board)*

| Product | Thickness [mm] | Density [kg/$m^3$] | Time to: | | | Flux to floor >20 kw/$m^2$ time [sec] | Moisture content [%] | Fire reaction class |
|---|---|---|---|---|---|---|---|---|
| | | | 1 MW Flashover [sec] | 600 kW Flashover [sec] | Flames exiting the door [sec] | | | |
| OSB | 11 | 643 | 177 | 168 | 189 | 186 | 5,88 | C |

Combining the above information regarding smoke generation and the quick reaction of OSB to reach Flashover led firemen to choose OSB as the main choice of material to be used in flashover container fire trainings.

The experiment is carried out using a container (single-room compartment) as shown in Figure 3. The container has the following dimensions: 12 m × 2.2 m × 2.6 m. A single ventilation opening was available, namely the front door of the container which remained open during the experiment.

Some of the results are presented in Figure 2. A thermal-vision camera has also been used in order to measure the temperature at the walls of the container and to validate the measurements values taken from the thermocouples.

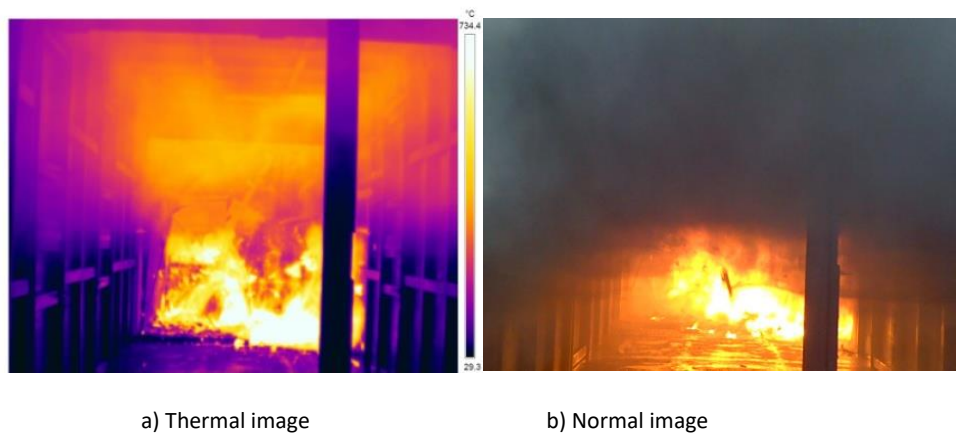

a) Thermal image　　　　　　　　b) Normal image

*Figure 2 - Post-flashover fire in under-ventilated flaming condition experiments*

Figure 2 presents images related to the flashover moment (in situ experiment) – all OSB combustible parts in the fire container are ignited. As one can observe from the thermal image, the temperature values greater than 650 °C are taken from the smoke and hot gas upper layer of the container and the temperature values greater than 700 °C are taken from the fuel surface area. This result is in complete agreement with the first physical characteristic of the flashover phenomena.

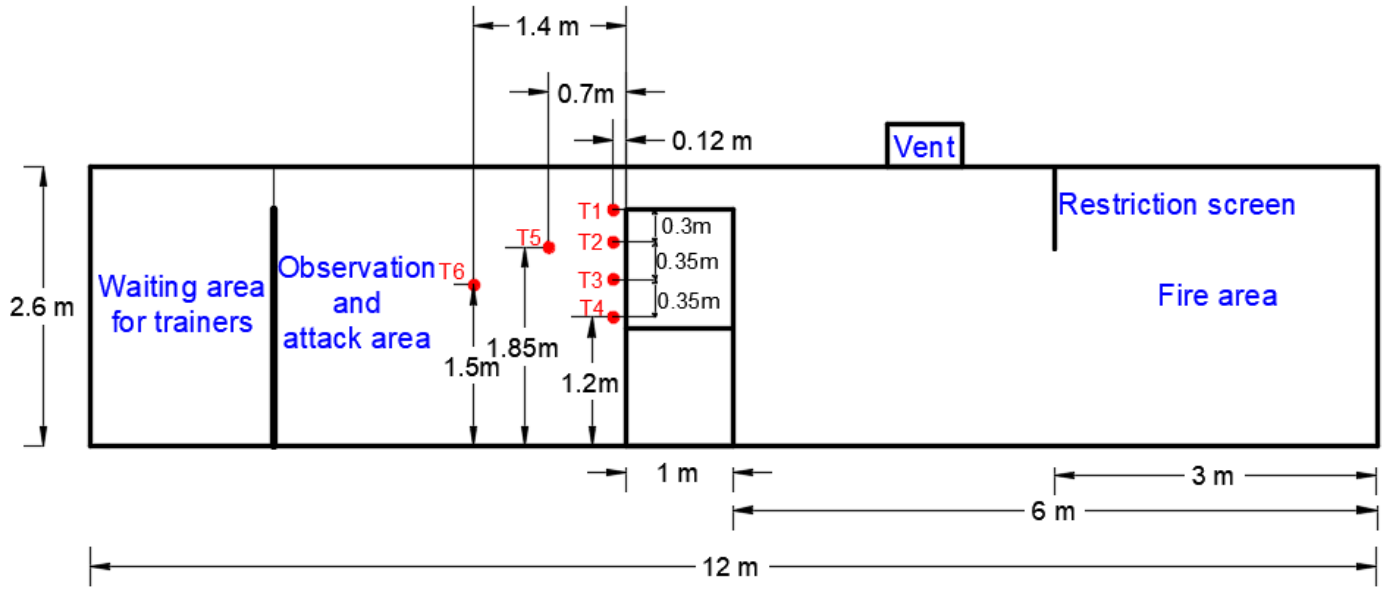

*Figure 3 - The right-side view scheme of arrangement (instrumentation) of the flashover container*

Parts of the walls and the ceiling of the container were furnished with oriented strand boards (see Figure 5). The fire source has been a wooden crib, made of 36 pieces of wood strips 2.5 cm × 2.5 cm× 30 cm (see Figure 4), on which has been poured 500 ml ethanol shortly before ignition. The fire bed was situated at 1.2 m below the ceiling (see Figure 4 and Figure 5).

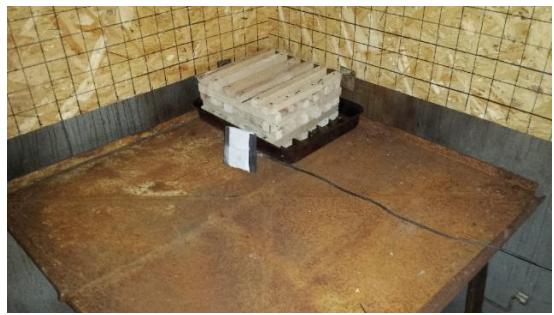

*Figure 4 - Location of the ignition burner*

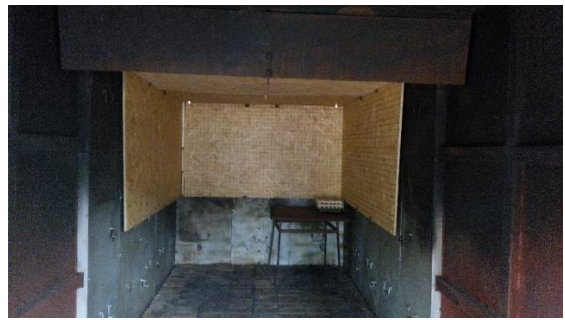

*Figure 5 - Location of the ignition burner*

The measurement devices (in front of observation and attack area) consisted in six built-in K-type thermocouples, fixed at key locations as shown in Figure 3, Figure 6 and Figure 7, connected to a data acquisition logger. Notice the similarities of the time-temperature plotting (Figure 8) with the idealized curve (Figure 1).

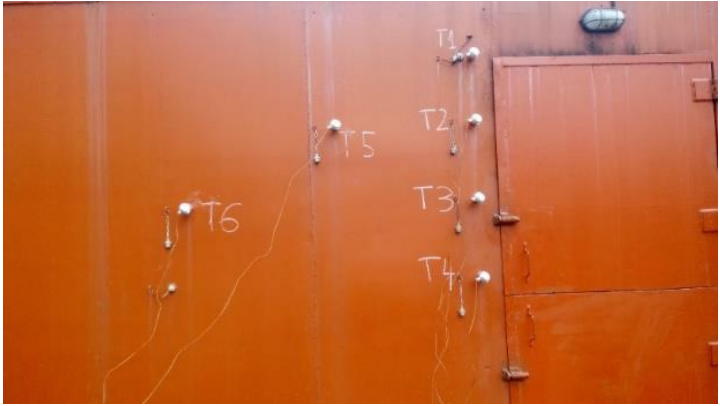

*Figure 6 - Thermocouples (exterior view)*

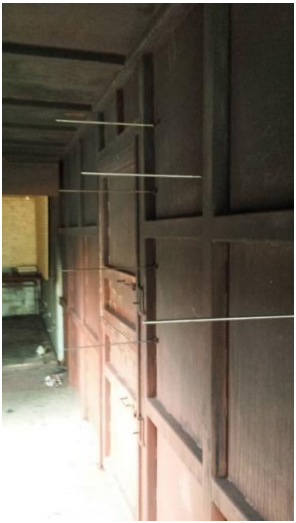

*Figure 7 - Position of the thermocouples (interior view)*

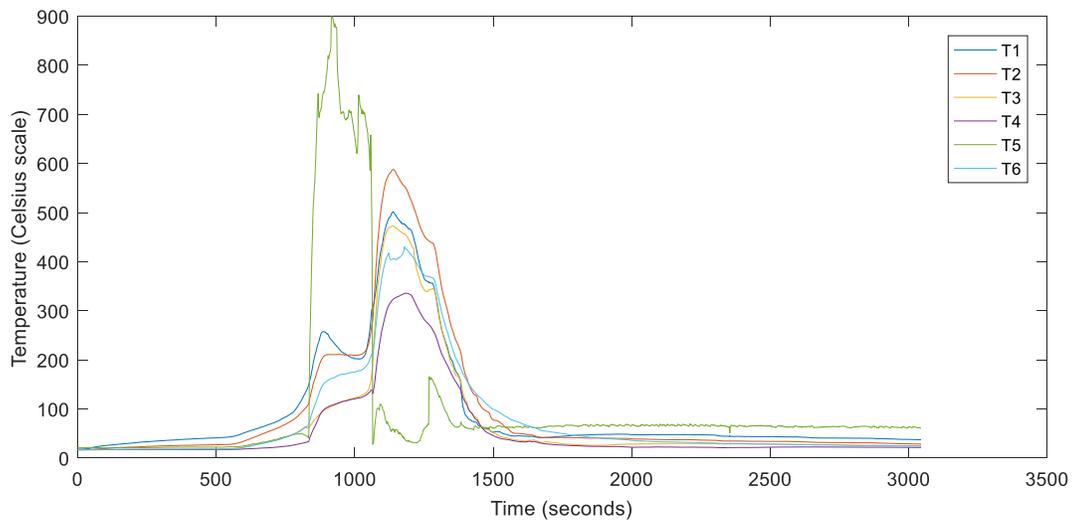

*Figure 8 - Time - temperature curves*

The temperature values outside the observation area (situated at the kneeled firefighter's head level) are large compared to the usual 450 - 500 °C, since the ventilation system was not used in this case – the ventilation helps controlling the fire dynamics. Only the thermocouple T4 records temperature values at the head level of the kneeled fire fighter (observation position).

In Table 2, Table 3, some basic statistical analysis of the temperatures (measured in degrees Celsius) allows us to quickly but roughly handle the experimental raw data.

*Table 2*

| Type of data/ Thermocouple number | T1 | T2 | T3 | T4 | T5 | T6 |
|---|---|---|---|---|---|---|
| max | 501.49 | 587.77 | 472.49 | 335.83 | 899.68 | 429.99 |
| min | 17.83 | 17.42 | 16.61 | 16.59 | 20.48 | 16.93 |
| Mean | 93.583 | 95.517 | 69.518 | 55.715 | 102.121 | 78.701 |
| Variance | 12483.933 | 17979.080 | 10972.692 | 5970.495 | 29815.258 | 10560.738 |

*Table 3 - Correlation Matrix*

| T1 | T2 | T3 | T4 | T5 | T6 |  |
|---|---|---|---|---|---|---|
| 1 | 0.986539 | 0.96682 | 0.965906 | 0.356321 | 0.964063 | **T1** |
| 0.986539 | 1 | 0.990619 | 0.991476 | 0.266062 | 0.990355 | **T2** |
| 0.96682 | 0.990619 | 1 | 0.988605 | 0.147797 | 0.97841 | **T3** |
| 0.965906 | 0.991476 | 0.988605 | 1 | 0.225958 | 0.992552 | **T4** |
| 0.356321 | 0.266062 | 0.147797 | 0.225958 | 1 | 0.269271 | **T5** |
| 0.964063 | 0.990355 | 0.97841 | 0.992552 | 0.269271 | 1 | **T6** |

# 3. Permutation entropy – Numerical Simulation

## 3.1. Theoretical background and remarks

*Shannon's entropy* (Shannon, 1948), widely accepted as a measure of disorder and uncertainty, is defined as $H(P) = -\sum_{i=1}^{n} p_i \log p_i$, where $P = (p_1, \ldots, p_n)$ is a finite probability distribution. It is positive and its maximum value is $H(U) = \log n$, where $U = \left(\frac{1}{n}, \ldots, \frac{1}{n}\right)$. Throughout the paper we use the convention $0 \cdot \log 0 = 0$.

The *normalized entropy* is $H(P)/\log n$. Adding impossible events to a set of probabilities does not affect its entropy, however it would clearly affect the normalized entropy, so we recommend to carefully interpret the results. We do not recommend omitting the so-called *forbidden patterns*, that is patterns (permutation/encoding sequences) that do not appear in the time series (with null frequencies). This would affect the normalized permutation entropy which becomes $h(j) = \frac{PE(j)}{\log(\#\{\pi: p_\pi > 0\})}$.

*The Kullback-Leibler divergence* (Kullback & Leibler, 1951) is defined by

$$D(P\|R) = \sum_{i=1}^{n} p_i (\log p_i - \log r_i),$$

where $P = (p_1, \ldots, p_n)$ and $R = (r_1, \ldots, r_n)$ are probability distributions.

*The Jensen-Shannon divergence (relative entropy)* is

$$JS(P\|R) = \frac{1}{2} D\left(P \left\| \frac{P+R}{2}\right.\right) + \frac{1}{2} D\left(R \left\| \frac{P+R}{2}\right.\right) = H\left(\frac{P+R}{2}\right) - \frac{H(P) + H(R)}{2}.$$

*The disequilibrium-based statistical complexity (LMC statistical complexity)* introduced in (López-Ruiz, Mancini, & Calbet, 1995) is defined as $C(P) = D(P) \frac{H(P)}{\log n}$, where $D(P)$, interpreted as disequilibrium, is the quadratic distance $D(P) = \sum_{i=1}^{n} (p_i - \frac{1}{n})^2$.

*The Jensen-Shannon statistical complexity* (Lamberti, Martin, Plastino, & Rosso, 2004), (Zunino, Soriano, & Rosso, 2012) is defined by $C^{(JS)}(P) = Q_{(JS)}(P) \frac{H(P)}{\log n}$, where the disequilibrium $Q_{(JS)}(P)$ is $Q_{(JS)}(P) = k \cdot JS(P\|U)$, where $k = (\max_P JS(P\|U))^{-1}$ is the normalizing constant and $U = \left(\frac{1}{n}, \ldots, \frac{1}{n}\right)$. For the computation of the normalizing constant, the maximum is attained for $P$ such that there exists $i, p_i = 1$.

For specific experimental requirements, one needs to calculate different kinds of entropies to make meaningful comparisons among various time series. For each entropy type we must study the corresponding LMC and Jensen-Shannon statistical complexity too.

### a) Extraction of the underlying probability distribution

*The permutation entropy* PE (Bandt & Pompe, 2002) as a complexity measure for time series is based on the appearance of ordinal patterns, that is on comparisons of neighboring values of time series, and it characterizes the diversity of the orderings in the time series, quantifying its complexity.

The basic principle of the PE-algorithm is as follows:

Let $T = (t_1, ..., t_n)$ be a time series with distinct values.

Step 1. Every $j$-tuple $(t_i, ..., t_{i+j-1})$, $i = 1, ..., n-j+1$ ($j$ is the embedding dimension, the window length) corresponds to a symbolic representation which has at most $j!$ different states. The increasing rearranging of the components of each $j$-tuple $(t_i, ..., t_{i+j-1})$ as $(t_{i+r_1-1}, ..., t_{i+r_j-1})$ yields a unique permutation of order $j$ denoted by $\pi = (r_1, ..., r_j)$, an encoding pattern that describes the up-and-downs in the considered $j$-tuple.

Step 2. The absolute frequency of this permutation (the number of $j$-tuples which are associated to this permutation) is

$$k_\pi \equiv \#\{i: i \leq n - (j-1), (t_i, ..., t_{i+j-1}) \text{ is of type } \pi\}.$$

These values have the sum equal to the number of all consecutive $j$-tuples, that is $n - (j-1)$.

Step 3. The *permutation entropy of order $j$* is defined as $PE(j) \equiv -\sum_\pi p_\pi \log p_\pi$, where $p_\pi = \frac{k_\pi}{n-(j-1)}$ is the relative frequency.

Simple numerical examples may help clarify the concepts throughout this section.

**Example** For the 5-tuple (2.3, 1, 3.1, 1.1, 5.2) the corresponding permutation (encoding) is (2, 4, 1, 3, 5).

By increasing the embedding dimension $j$ one can check the consistency of the results. The main issue in every entropy-based approach is choosing the appropriate embedding dimension $j$.

**Remark** It is known that $0 \leq PE(j) \leq \log j!$. The lower bound is attained for an increasing or decreasing sequence (time series), and the upper bound for randomly distributed sequences where all $j!$ possible permutations are equiprobable (so called white noise, in signal processing). For the randomness, we have limitations due to the necessary condition $j! | (n - (j-1))$, since the factorial increases fast and one requires a bigger $n$. So, as guideline for choosing the embedding dimension, the value of the permutation entropy remains relevant for $j$ such that $n \gg j!$, therefore one applies the PE-algorithm only for sufficiently small $j$, avoiding the big values of the factorial. Usually one takes $j$=3,4,5,6,7. See also (Kulp & Zunino, 2014).

In addition, we note that the time series obtained during fire experiments cannot have only increasing (or decreasing) $j$-tuples, and consequently the permutation entropy is strictly positive (the most commonly encountered tuples are, for all $j$, the increasing and the decreasing ones).

**Remark** In an ideal fire experiment, the permutation entropy $PE(j)$ becomes minimal in the special case where the temperatures have the evolution described by a curve with exactly one extreme point and strict monotonicity before/after that (as the standard curves in Section 1: most of the $j$-tuples are increasing or decreasing). The divergence

$$D^{(PE)}(P\|U) = \sum_{i=1}^{n} p_i (\log p_i - \log \frac{1}{j!}) = \log j! - PE(j)$$

would then be maximal.

Since we have not found in the literature any remark about the lower bound of the permutation entropy for fire experiments, theoretically determined or inferred from experimental data, we propose the next open problem.

**Open Problem** Find the greatest lower bound (threshold) $a(k)$ such that, if the probability distribution $P = (p_1, \ldots, p_n)$ has at least $k$ nonzero and equal components, $\geq a(k)$, then the Shannon entropy $H(P) = -\sum_{i=1}^{n} p_i \log p_i$ attains its minimum when $n - k$ components are zero. The case $k = n$ is degenerated, the required minimum equals then the maximum. Is this true in the more general setting "at least $k$ nonzero components", cancelling the equality condition and finding the greatest lower bound of the sum of $k$ nonzero components $(b(k))$? Is it true that $b(k) = ka(k)$?

In (Bandt & Pompe, 2002) the measured values of the time series are considered distinct. The authors neglect equalities and propose to break them by adding small random perturbations (random noise) to the original series. However, the equal values might characterize a specific stage of the phenomenon. In other words, if we ignore or eliminate the equal values, we do not always accurately describe the complexity of the system.

Another known solution is to rank the equalities according to their *order of emergence* (to rank the equalities with their sequential orders), a method recommended in the literature, see for instance (Cao, Tung, Gao, Protopopescu, & Hively, 2004) and (Duan, Wang, & Zhang, 2019).

**Example** For the 5-tuple (2.3, 1, 3.1, 1, 5.2) the corresponding permutation (encoding) is (2, 4, 1, 3, 5).

This manner to adjust the computation can be used to analyze the statistical structure of fire experimental data, since the temperatures are sometimes equal due to the thermal inertia of the thermocouples or to too frequent measuring. At our best knowledge there is no method to estimate a relevant time interval to measure the temperatures during an experiment, any attempt would depend on the devices one uses and on the fuel.

**Remark** In order to interpret our results correctly, we emphasize some limitations of the computation due to the artificial ordering of the equal temperatures: if $(t_i, \ldots, t_{i+j-1})$ is of type $\sigma = (1, 2, \ldots, j)$ then the $j$-tuple is monotonically increasing, $t_i \leq \cdots \leq t_{i+j-1}$, and if the $j$-tuple is of type $\tau = (j, j-1, \ldots, 1)$ then the temperatures are strictly decreasing, $t_i > \cdots > t_{i+j-1}$. Computing $p_\sigma$ refers to the growing phase of the fire, but $p_\tau$ is only partially analyzing the decreasing temperatures. One cannot expect that the probability distribution has equal (or almost equal) terms, that is the entropy does not approach its maximum: all our experimental data show a $p_\sigma$ greater than the rest of the probabilities.

Weighted permutation entropy WPE (Fadlallah, Chen, Keil, & Príncipe, 2013)

There are many ways to associate a probability distribution to the given data set $T = (t_1, \ldots, t_n)$. The way PE is defined shows that no other information besides the order structure is retained. The importance of changes in the amplitude of signals for distinguishing different states is emphasized in the literature, so as to differentiate between distinct tuples of a certain encoding pattern. Hence, the relevance of the permutation entropy can be improved if the variance of each $j$-tuple $(t_i, \ldots, t_{i+j-1})$ is considered.

The dispersion (biased sample variance) is

$$w_i = \frac{1}{j}\sum_{k=i}^{i+j-1}(t_k - \bar{t})^2, i \leq n - (j-1),$$

where $\bar{t}$ is the arithmetic mean of $t_i, \ldots, t_{i+j-1}$.

The weighted relative frequency which corresponds to an encoding pattern (permutation) $\pi$ is

$$p(\pi) = \frac{\sum_{i:(t_i,...,t_{i+j-1}) \text{ is of the type } \pi} w_i}{\sum_{i=1}^{n-j+1} w_i}.$$

By incorporating the amplitude information from the relative order structure, the *weighted permutation entropy* (WPE($j$)) is defined as $\text{WPE}(j) \equiv -\sum_\pi p(\pi) \log p(\pi)$. (Fadlallah, Chen, Keil, & Príncipe, 2013), (Zhang & Shang, 2018). It weights differently $j$-tuples having the same ordinal pattern but different amplitude variations.

### b) Other variants

We also intend to test other algorithms used to extract the underlying probability distribution.

The *modified permutation entropy* (mPE) has been introduced in (Bian, 2012) as follows. For distinct temperatures, one applies the PE-algorithm. When equality occurs, the equal values are mapped onto the same symbol, which is the smallest time index of these equal values: if $t_{i+r_1-1} = t_{i+r_2-1}$, $r_1 < r_2$ then both temperatures will be represented by $r_1$ in the encoding symbol sequence (not a permutation anymore, as for PE). The obtained probability distribution for these symbol sequences (encodings) is used to compute Shannon entropy and the result is called the *modified permutation entropy* (mPE($j$)).

**Remark** When a lot of equalities occur, the mPE method is expected to perform better since it characterizes more states than the PE method. However, when the number of equalities is small in comparison to the amount of measurements, the values of mPE would be almost equal to those of PE, as it will be visible in the analysis of our experimental data. This can be mathematically explained as follows, by Fadeev's postulate (Fadeev, 1957):

$$H(tp_1, (1-t)p_1, p_2, ..., p_n) = H(p_1, p_2, ..., p_n) + p_1 H(t, (1-t)) \text{ for } t \in [0,1].$$

For every permutation $\pi$ used by the PE-algorithm, which corresponds to a $j$-tuple containing an inequality $t_i \leq t_{i+1}$, we have two encodings in the mPE-algorithm, one for the $j$-tuples containing $t_i < t_{i+1}$, respectively the other one for $t_i = t_{i+1}$, and their relative frequencies have the sum $p(\pi)$, so $\text{mPE}(j) \geq \text{PE}(j)$.

Sometimes, the fire experimental data contains consecutively measured equal values of the temperature. These equalities can be explained by the resolution that might be too high (measurements every second) due to the thermal inertia of the devices. One can try to avoid this drawback by selecting a coarse resolution (a different time scale), however the best interval length is yet to be established in order not to lose important information.

**Remark** The mPE-algorithm deals with at most 13 encoding sequences for $j=3$, as for $j=4$ there are 75. Finding a general formula to compute the number of the symbol sequences for a general embedding dimension $j$ can be stated as an interesting combinatorial question (the answer is greater than $j!$).

**Example** (2.3, 1, 3.1, 1, 5.2) → (2, 2, 1, 3, 5)

**Definition** We introduce the *weighted modified permutation entropy* (WmPE) by the following computational algorithm: the $j$-tuples are encoded according to the mPE method, followed by the computation of WmPE($j$) using weights computed from the variances, as described at the WPE algorithm above.

WmPE($j$) is compensating the loss of the information carried by mPE($j$) for small numbers of equalities and it extends the concept of mPE($j$) while keeping the same Shannon's entropy expression, as WPE($j$) extends the definition of PE($j$). The permutation/encoding type entropies (i.e. PE, mPE, WPE, WmPE) depend on the considered embedding dimension $j$ and one still must determine an appropriate value of it via meaningful comparisons.

It is worth to be noted that for computing the weighted modified permutation entropies the constant valued pattern brings no additional amplitude information (its variance is null).

*Two-length algorithm* (Watt & Politi, 2019)

**Step 1**. Given the $j$-tuple $T = (t_1, \ldots, t_j)$, we start encoding the last $k \leq j$ elements $(t_{j-k+1}, \ldots, t_j)$ according to the ordinal position of each element, that is every $t_s$ is replaced by a symbol which indicates the position occupied by $t_s$ within the increasing rearranging of the considered $k$-tuple.

**Example** $(3.1, 5.2, 1.1) \rightarrow (2, 3, 1)$ for $k = 3$.

**Remark** If we compare this step with the standard implementation of the permutation entropy, according to the PE-algorithm (Bandt & Pompe, 2002), we see that the algorithm suggested in (Watt & Politi, 2019) is providing a permutation which is the inverse of the one provided by the PE-algorithm. The PE-algorithm would encode $(3.1, 5.2, 1.1)$ by the permutation $(3, 1, 2)$.

**Step 2.** Next, we proceed by encoding each previous element $t_m$ up to $m = 1$ according to the symbol provided by Step 1 applied to the $k$-tuple $(t_m, \ldots, t_{m+k-1})$.

**Example** $(3.4, 2.3, 3.1, 5.2, 1.1) \rightarrow (3, 1, 2, 3, 1)$ for $k = 3$ and $j=5$.

Given the pair $(k, j)$ of values, the number of symbolic (encoding) sequences of length $j$ is $k! \, k^{j-k}$, a number which can be much smaller than $j!$, so this algorithm is faster, it involves a simplified computation and sometimes it makes the results more relevant for big values of $j$.

Notice that the encoding step is not telling how to deal with equal values, so in that case we will use the technique described for the permutation entropy PE: we consider the chronological order (encoding of type 1). Alternatively, one can apply the above two-length algorithm and map the equal values with identical symbols (encoding of type 2; we call this *the modified two-length algorithm*). These algorithms lead, after computing the relative frequencies of the encoding sequences, to two different entropies: *the two-length permutation entropy* (TLPE$(k,j)$) and *the modified two-length permutation entropy* (mTLPE$(k,j)$).

**Example** a) Encoding of type 1: $(3.1, 3.1, 3.1, 1, 3.1) \rightarrow (1, 2, 2, 1, 3)$ for $k = 3$ and $j=5$.

b) Encoding of type 2: $(3.1, 3.1, 3.1, 1, 3.1) \rightarrow (1, 2, 2, 1, 2)$ for $k = 3$ and $j=5$.

The weighted entropies WTLPE$(k,j)$ and WmTLPE$(k,j)$ can be now easily introduced: in order to compute them we follow the two-length algorithm (respectively the modified two-length algorithm) to encode the tuples, we use an encoding of type 1 (respectively of type 2) for the equal values and compute the entropy using Shannon's definition with the probability distribution defined as above, for WPE($j$).

The algorithm for the modified permutation type entropies returns a big number of encoding sequences, a fact that reduces the possibility to obtain significant results for mTLPE$(k,j)$ and WmTLPE$(k,j)$ unless the time series are very long: as we mentioned, the size of the time series under consideration must be much bigger than the number of encoding sequences, an aspect emphasized already for the permutation entropy PE($j$). This is the reason why we present in the next section only the entropies TLPE(3,5) and WTLPE(3,5).

In the next section we apply these techniques and observe their capability to discern the changes in the complexity of the experimental data.

## 3.2. Raw data analysis

The raw data set under consideration consists of measured temperatures during a compartment fire: six thermocouples T1, …, T6 measure the temperatures every second during the experiment. Hence, we get six time-series consisting of 3046 entries (data points) and we aim to obtain a better understanding of these results by modeling the time series using information theory, and to evaluate the performance of the discussed entropies.

The most frequent $j$-tuples are the increasing and the decreasing ones for $j$=3,4,5. The (common) rare patterns among the 3-tuples are (1,3,2) and (2,1,3) (see the picture below), that is when $t_i \leq t_{i+2} < t_{i+1}$ or $t_{i+1} < t_i \leq t_{i+2}$. In other words, except in the case of the monotonically increasing 3-tuples, the 3-tuples usually have the initial temperature higher than the last one, a fact which agrees with our intuition. See in Figure 9, Figure 10, Figure 11 these patterns which *rarely* appear in the evolution of the temperature (encoding of type 1).

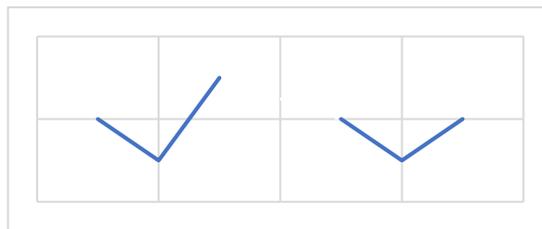

Figure 9 - Patterns for the 3-tuple (2,1,3)

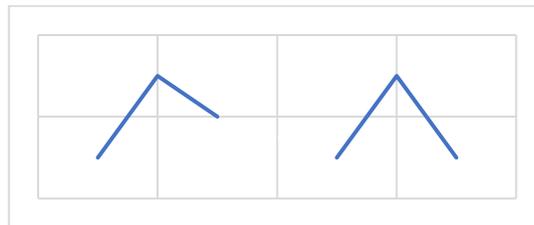

Figure 10 - Patterns for the 3-tuple (1,3,2)

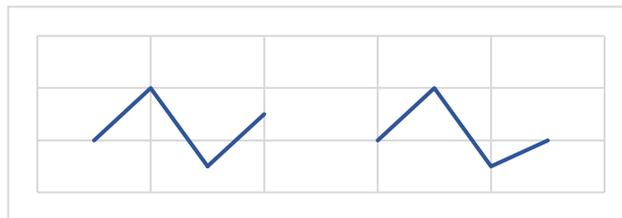

Figure 11 - Patterns for the 4-tuple (3,1,4,2)

We consider that further observations are needed to check the claimed patterns of the rare $j$-tuples for fire experiments and their correlation to the permutation entropy. Naturally, the 4 and 5-tuples which contain rare 3-tuples are also rare (with smaller frequency), although our experimental data shows that the 4-tuple with the smallest frequency (3,1,4,2) does not contain any of the rare 3-tuples and its frequency is also less correlated to the permutation entropy (see Table 5).

In Table 4, Table 5, and Table 6 one can see that there exists a high positive correlation between the permutation entropy and the relative frequencies of the rare patterns.

Table 4 - Relative frequencies of rare 3-tuples

|    | (1,3,2)  | (2,1,3)  | PE(3)    |
|----|----------|----------|----------|
| T6 | 0.033837 | 0.037779 | 1.369525 |
| T2 | 0.039750 | 0.039422 | 1.384373 |
| T3 | 0.053876 | 0.058147 | 1.417392 |
| T4 | 0.055191 | 0.058804 | 1.444837 |
| T1 | 0.068660 | 0.069645 | 1.478165 |
| T5 | 0.113666 | 0.106110 | 1.686625 |

Table 5 - Relative frequencies of rare 4-tuples

|    | (3,1,4,2) | PE(4)    |
|----|-----------|----------|
| T6 | 0.002958  | 2.022171 |
| T2 | 0.002300  | 2.053314 |
| T3 | 0.003286  | 2.139724 |
| T4 | 0.003286  | 2.215603 |
| T1 | 0.007558  | 2.319293 |
| T5 | 0.012488  | 2.863961 |

Table 6 - Relative frequencies of rare 5-tuples

|    | (3,1,5,4,2) | (4,2,1,5,3) | PE(5)    |
|----|-------------|-------------|----------|
| T6 | 0.000000    | 0.000000    | 2.659253 |
| T2 | 0.000000    | 0.000000    | 2.685129 |
| T3 | 0.000000    | 0.000000    | 2.852491 |
| T4 | 0.000000    | 0.000000    | 2.990761 |
| T1 | 0.001315    | 0.000657    | 3.183472 |
| T5 | 0.001972    | 0.002630    | 4.148164 |

It is worth mentioning here that we did not observe any forbidden pattern being common to all the time series (thermocouples) for $j$=3,4,5.

We aim to identify the behavior and those physical properties of the combustion phenomena which are captured by different permutation entropies of the collected temperature measurements, and the main difficulty is to establish the appropriate entropy formulas to be used for the research of fire data.

The values of the above discussed permutation/encoding type entropies are given in the table below.

Table 7 - Permutation/encoding entropies

|        | T1          | T2          | T3          | T4          | T5          | T6          |
|--------|-------------|-------------|-------------|-------------|-------------|-------------|
| PE(3)  | 1,478164565 | 1,384373014 | 1,417392093 | 1,444837353 | 1,686624662 | 1,369525342 |
| WPE(3) | 0,712938044 | 0,603970981 | 0,622302478 | 0,649001671 | 0,990873687 | 0,706862416 |
| mPE(3) | 1,826103448 | 1,88768009 | 2,072776342 | 2,120708887 | 1,839524345 | 1,920941417 |
| WmPE(3) | 0,714645861 | 0,604384072 | 0,622861236 | 0,64964279 | 0,991070005 | 0,707286963 |
| PE(4) | 2,319293204 | 2,053313765 | 2,139724467 | 2,215603332 | 2,863960958 | 2,022171411 |
| WPE(4) | 0,746800657 | 0,604841624 | 0,627010902 | 0,682167991 | 1,46127927 | 0,761193063 |
| mPE(4) | 2,859018051 | 2,716119454 | 3,035415313 | 3,124622138 | 3,116983367 | 2,751657353 |
| WmPE(4) | 0,74894709 | 0,605374684 | 0,628011131 | 0,683109795 | 1,4615796 | 0,761867376 |
| TLPE(2,5) | 2,239641492 | 2,015247958 | 2,107322513 | 2,165408031 | 2,637116668 | 1,978177599 |
| WTLPE(2,5) | 0,783540463 | 0,605354367 | 0,630572943 | 0,713694735 | 1,692075211 | 0,787154348 |
| TLPE(3,5) | 2,764664174 | 2,353814472 | 2,533044226 | 2,666280079 | 3,443692711 | 2,327780461 |
| WTLPE(3,5) | 0,786531619 | 0,605449058 | 0,630820756 | 0,711956588 | 1,814913577 | 0,792826432 |
| mTLPE(2,5) | 2,803993256 | 2,846190089 | 3,24587624 | 3,344655217 | 2,871489655 | 2,90532114 |
| WmTLPE(2,5) | 0,785924196 | 0,605881758 | 0,632186985 | 0,714908577 | 1,69237014 | 0,788188839 |
| PE(5) | 3,183471839 | 2,68512949 | 2,85249149 | 2,990761238 | 4,148164086 | 2,659253034 |
| WPE(5) | 0,792732616 | 0,605628856 | 0,631543081 | 0,714894073 | 1,882221649 | 0,798759327 |

Quick comparisons are provided in the next figures.

We observe that the use of the weighting algorithm (computing the probability distribution via the variance) returns very similar plottings of the entropies, sometimes almost identical, for different embeddings.

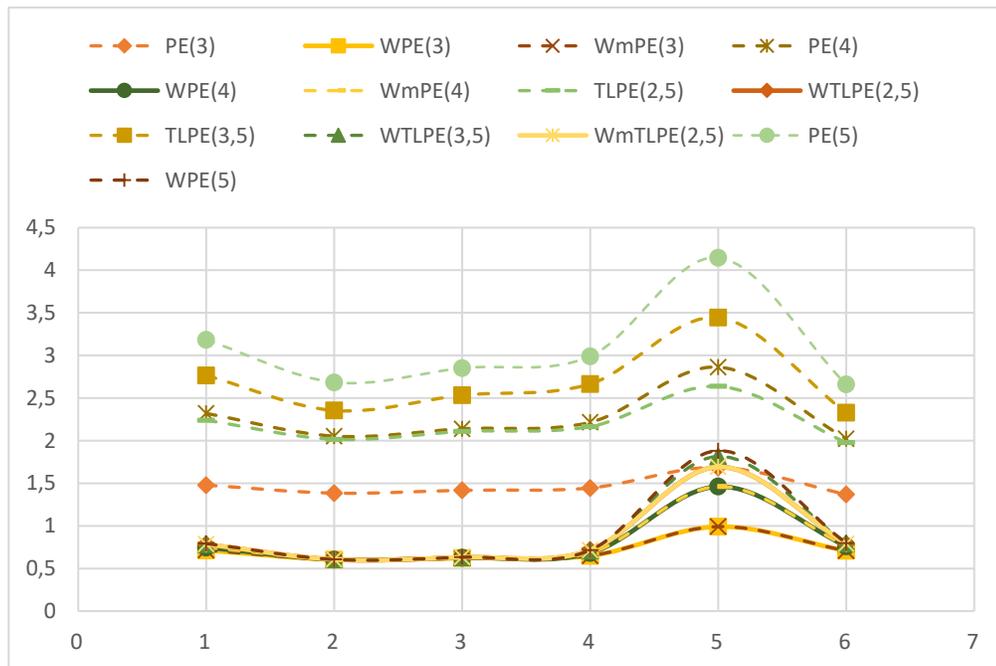

*Figure 12 - Permutation/encoding entropies (a)*

The modified entropies have their values closer to the original entropies if less equalities occur in the time series, a fact which is visible in the thermocouple T5.

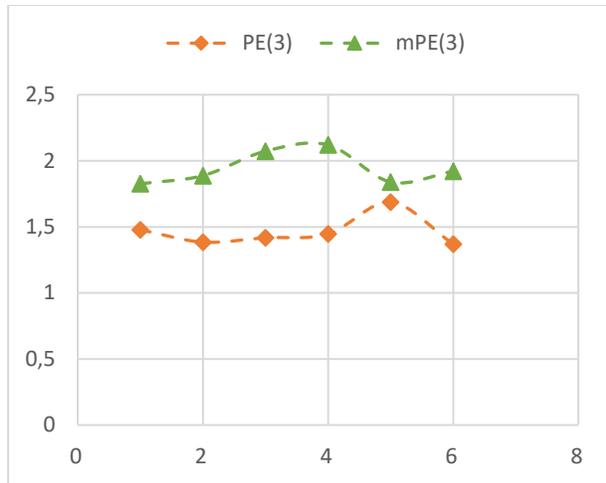
Figure 13 - Permutation/encoding entropies (b)

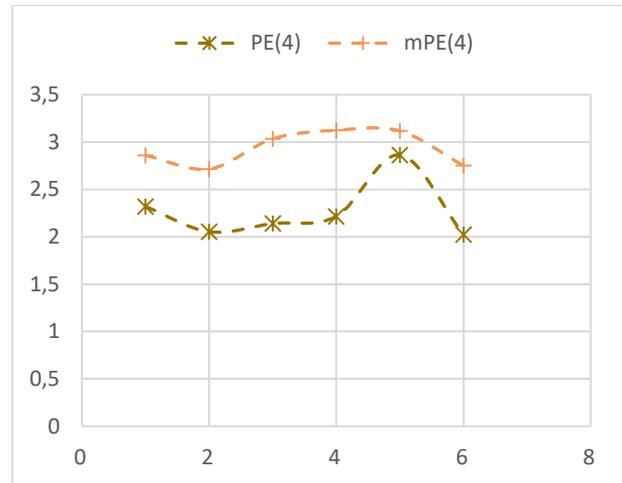
Figure 14 - Permutation/encoding entropies (c)

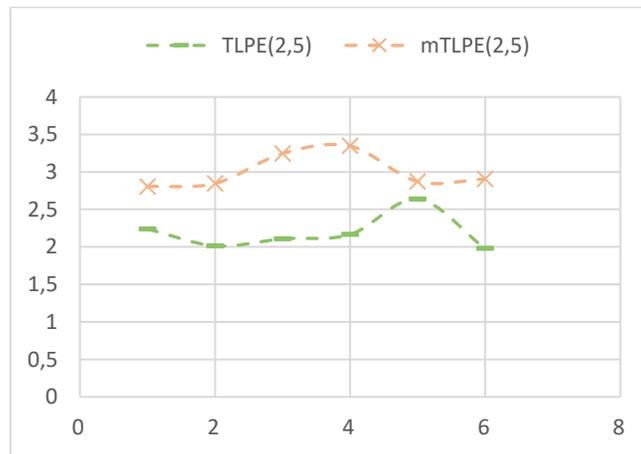
Figure 15 - Permutation/encoding entropies (d)

Table 8 - Number of constant j-tuples

|    | j=2 | j=3 | j=4 | j=5 |
|----|-----|-----|-----|-----|
| T1 | 141 | 18  | 5   | 2   |
| T2 | 360 | 92  | 20  | 6   |
| T3 | 486 | 148 | 42  | 11  |
| T4 | 546 | 170 | 58  | 19  |
| T5 | 42  | 3   | 0   | 0   |
| T6 | 462 | 158 | 57  | 17  |

In Figure 16 we plot these entropies to visualize our conclusions. On the *x*-axis we have all the 16 permutation/encoding entropies used (ordered as in Table 7) and on the *y*-axis the values at each thermocouple. All entropies lead us to the same conclusion: the values at the thermocouple T5 are much different than the others

(indicating that something is going on there), except the modified entropies mPE and mTLPE. Note that mPE and mTLPE are not pointing out the turbulence at T5, in the sense that they are not providing higher values (all the other entropies exhibit a higher degree of randomness for T5, visible also on the temperature-time plotting). The mathematical explanations consists in the number of equalities that is not the same at all thermocouples: in fact at T5 it is much smaller, and the modified entropy becomes significantly closer to its corresponding entropy at T5, while for other thermocouples this is not true. When we apply the weighting algorithm (that is when we compute the WmPE and WmTLPE), the higher variances are compensating the lack of equalities.

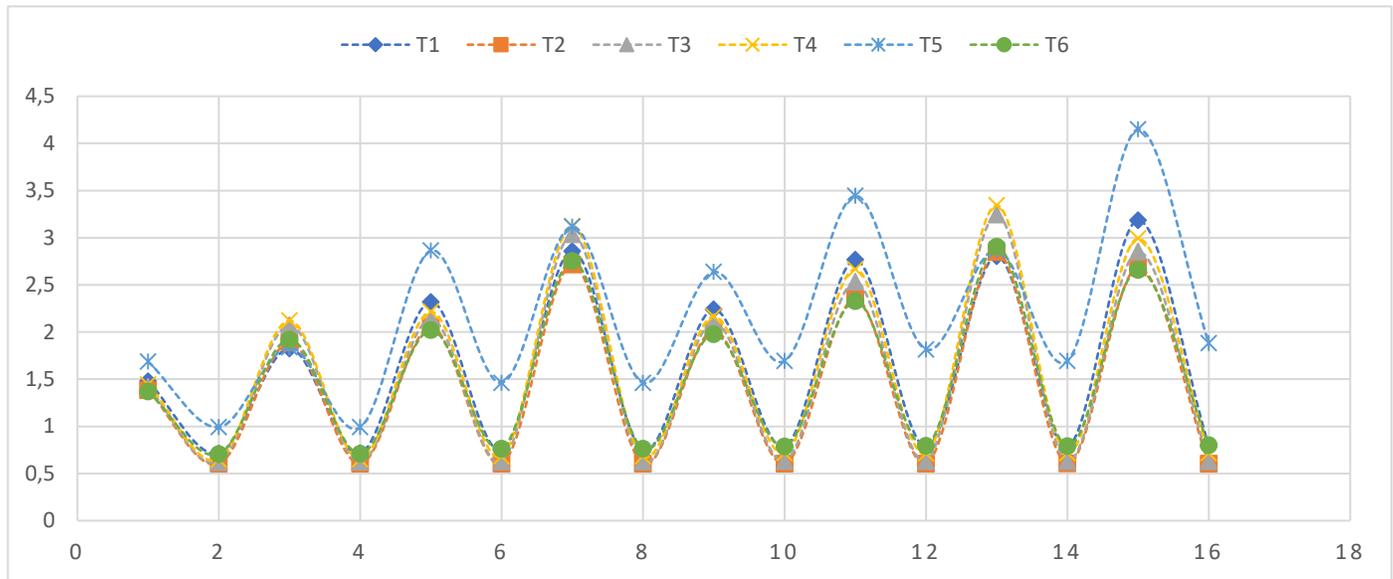

Figure 16 - Permutation/encoding entropies

We conclude that the modified entropies are not suitable for our analysis, probably due to the relatively small numbers of equalities which occur. All the other entropies prove themselves relevant for detecting unusual measurements among the thermocouples. Moreover, regardless the embedding dimensions and other length parameters, we obtained the following ordering of the entropies PE and TLPE: $T2 < T6 < T3 < T4 < T1 < T5$.

In what follows we analyze the dynamical behavior of the temperature in the compartment fire from the viewpoint of statistical complexity. The novelty of our approach consists in investigating the *LCM* and the *Jensen-Shannon disequilibrium-based statistical complexities* ($C(P)$ and $C^{(JS)}(P)$) using the techniques described above (that is successively plugging the entropies PE (and its variants WPE, mPE, WmPE) and TLPE (WTLPE, mTLPE, WmTLPE) in the $C(P)$ (respectively $C^{(JS)}(P)$) formula. See Table 9 and Table 10 for the statistical complexities established with the data gathered during our experiment. Each line contains the values of the statistical complexities obtained for the entropies listed in the first column.

*Table 9 - LMC Statistical Complexity*

| Entropy/Thermocouple | T1 | T2 | T3 | T4 | T5 | T6 |
|---|---|---|---|---|---|---|
| PE(3) | 0,098480327 | 0,11182792 | 0,112722303 | 0,105852504 | 0,035691422 | 0,112523668 |
| WPE(3) | 0,133841195 | 0,141819935 | 0,141609645 | 0,143290526 | 0,159586053 | 0,150657494 |
| mPE(3) | 0,113841465 | 0,113167742 | 0,085561077 | 0,07629502 | 0,083294861 | 0,112998742 |
| WmPE(3) | 0,118686871 | 0,12027297 | 0,120789687 | 0,122912715 | 0,146175862 | 0,130045196 |
| PE(4) | 0,108468333 | 0,121709585 | 0,120789664 | 0,111428927 | 0,035961085 | 0,122669959 |
| WPE(4) | 0,107288684 | 0,103787337 | 0,10515609 | 0,111085411 | 0,125169491 | 0,119362094 |
| mPE(4) | 0,092249122 | 0,099679986 | 0,07705676 | 0,071601417 | 0,044571344 | 0,100083535 |
| WmPE(4) | 0,084101303 | 0,080429272 | 0,081633416 | 0,086353979 | 0,101744924 | 0,092931336 |
| TLPE(2,5) | 0,085522956 | 0,109797529 | 0,100230538 | 0,088337083 | 0,018433049 | 0,112773046 |
| WTLPE(2,5) | 0,121090377 | 0,114460606 | 0,116527036 | 0,126945613 | 0,10149733 | 0,133741087 |
| TLPE(3,5) | 0,096105423 | 0,111132875 | 0,105774865 | 0,096443436 | 0,031150893 | 0,114168653 |
| WTLPE(3,5) | 0,09314309 | 0,086245209 | 0,087980545 | 0,095871253 | 0,092985728 | 0,102366916 |
| mTLPE(2,5) | 0,077568155 | 0,096187535 | 0,07442691 | 0,069712758 | 0,04151491 | 0,102565167 |
| WmTLPE(2,5) | 0,08558695 | 0,079187905 | 0,080897911 | 0,088375381 | 0,083362826 | 0,093476858 |
| PE(5) | 0,092095916 | 0,102359979 | 0,098311687 | 0,090917893 | 0,024424881 | 0,102509416 |
| WPE(5) | 0,079846391 | 0,07316938 | 0,074733145 | 0,081729172 | 0,083705499 | 0,087619931 |

In Figure 17 we plot these values: on the *x*-axis the LMC statistical complexities corresponding to the entropies (ordered as in Table 9) are situated and on the *y*-axis the values at each thermocouple. Notice that at T5, regardless of the embedding dimension *j*, the entropies PE, mPE, TLPE, mTLPE lead to LMC statistical complexities which are much different than those obtained at T1, T2, T3, T4, T6, this being a sign that the values of this time series show an unusual evolution of the phenomenon.

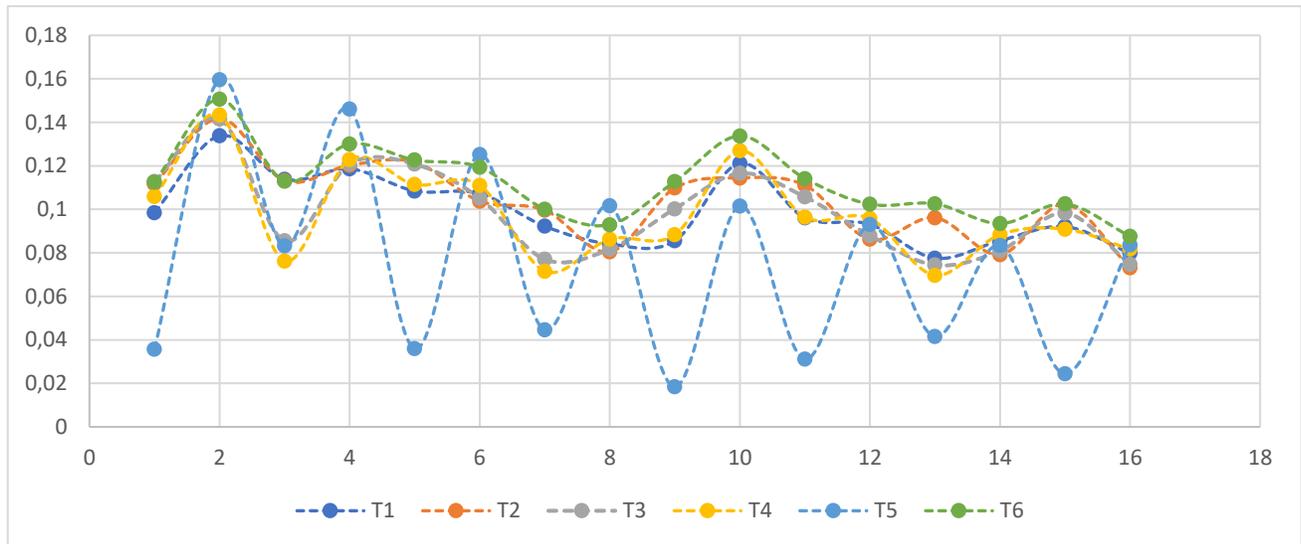

*Figure 17 - LCM Statistical Complexity*

*Table 10 - Jensen-Shannon Statistical Complexity*

| Entropy/Thermocouple | T1 | T2 | T3 | T4 | T5 | T6 |
|---|---|---|---|---|---|---|
| PE(3) | 0,137698439 | 0,174836879 | 0,158740583 | 0,150014621 | 0,052909618 | 0,181045251 |
| WPE(3) | 0,270309641 | 0,244721991 | 0,249124394 | 0,253586368 | 0,259463393 | 0,258088261 |
| mPE(3) | 0,22012102 | 0,198114055 | 0,160315074 | 0,150532609 | 0,253497176 | 0,191392104 |
| WmPE(3) | 0,229452362 | 0,200679875 | 0,205495905 | 0,211433453 | 0,26703573 | 0,221476861 |
| PE(4) | 0,200425629 | 0,244172343 | 0,230924921 | 0,224528775 | 0,099538786 | 0,247698738 |
| WPE(4) | 0,204487177 | 0,172601314 | 0,177573568 | 0,188300932 | 0,29562408 | 0,201936767 |
| mPE(4) | 0,253428853 | 0,258255454 | 0,245121261 | 0,236636663 | 0,300637931 | 0,262199824 |
| WmPE(4) | 0,162329939 | 0,134706374 | 0,139009219 | 0,148928619 | 0,277625777 | 0,162297511 |
| TLPE(2,5) | 0,15987331 | 0,207575358 | 0,190988681 | 0,184078043 | 0,05392225 | 0,215751399 |
| WTLPE(2,5) | 0,229118668 | 0,190635922 | 0,19619801 | 0,211009872 | 0,298379594 | 0,223302132 |
| TLPE(3,5) | 0,22924386 | 0,270715277 | 0,257956531 | 0,248396078 | 0,153914156 | 0,269176737 |
| WTLPE(3,5) | 0,179416657 | 0,144128944 | 0,149033342 | 0,163688372 | 0,323670651 | 0,177427226 |
| mTLPE(2,5) | 0,296679906 | 0,248178638 | 0,209569777 | 0,194436319 | 0,357031883 | 0,235344152 |
| WmTLPE(2,5) | 0,166234538 | 0,132751162 | 0,137532786 | 0,151879152 | 0,316268153 | 0,164653848 |
| PE(5) | 0,239599602 | 0,282725584 | 0,278112097 | 0,271652231 | 0,152549818 | 0,282377642 |
| WPE(5) | 0,155671387 | 0,123039451 | 0,12749253 | 0,14147375 | 0,312232082 | 0,154887173 |

In Figure 18, on the *x*-axis we put the Jensen-Shannon statistical complexities corresponding to the entropies (ordered as in Table 10) and on the *y*-axis the values at each thermocouple. Summarizing, at T5, all the entropies we tested lead to Jensen-Shannon statistical complexities which are different than those obtained at T1, T2, T3, T4, T6, this not only being a sign of an unusual evolution of the phenomenon at T5, but also enabling us to conclude that the Jensen-Shannon formula is more helpful in the analysis of the compartment fire behavior (we encourage the specialists to test it on more experimental setups of compartment fire).

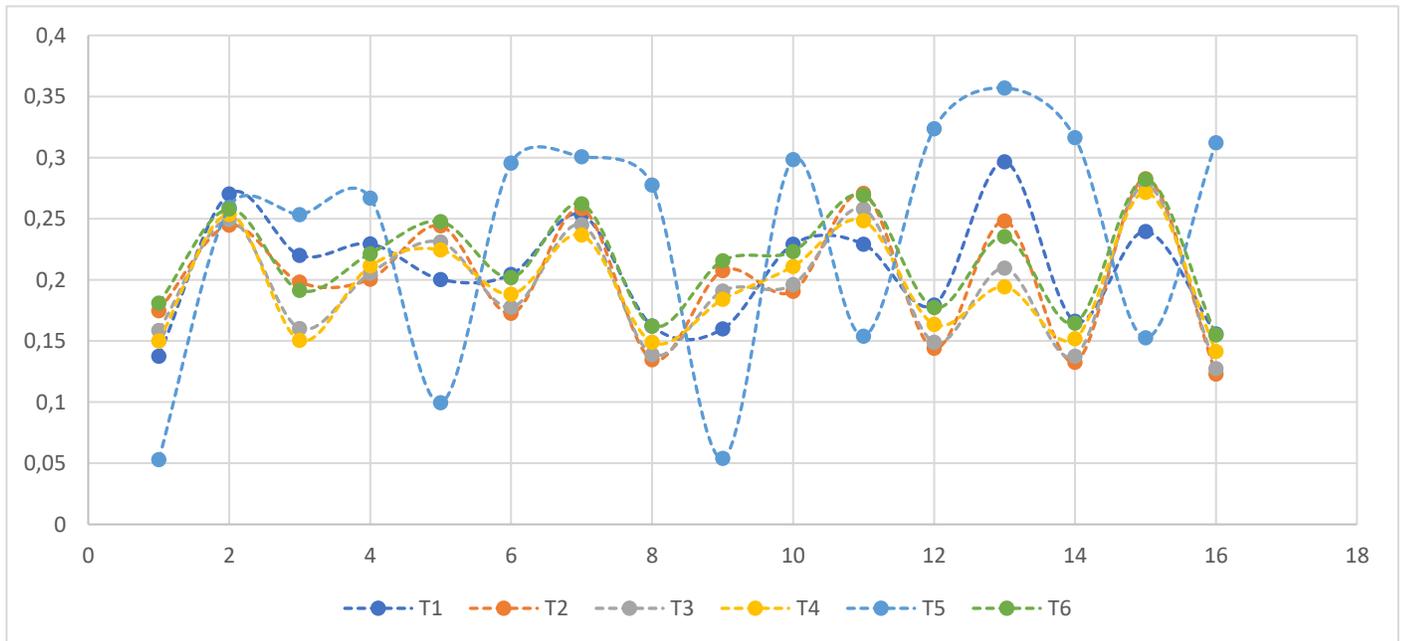

*Figure 18 - Jensen-Shannon Statistical Complexity*

See below other comparisons of the gathered data. From their analysis (especially at the thermocouple T5), we maintain our recommendation, for further studies, to use the weighted entropies (and, whenever possible, the

(weighted) modified ones), which seem to shed more light on the Jensen-Shannon statistical complexity and provide sharper tools to establish the disequilibrium and turbulence-related characteristics.

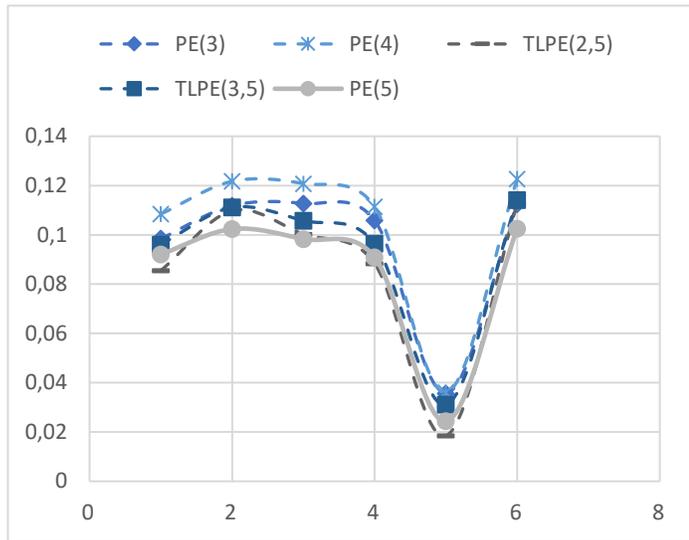

*Figure 19 - LMC Statistical Complexity (a)*

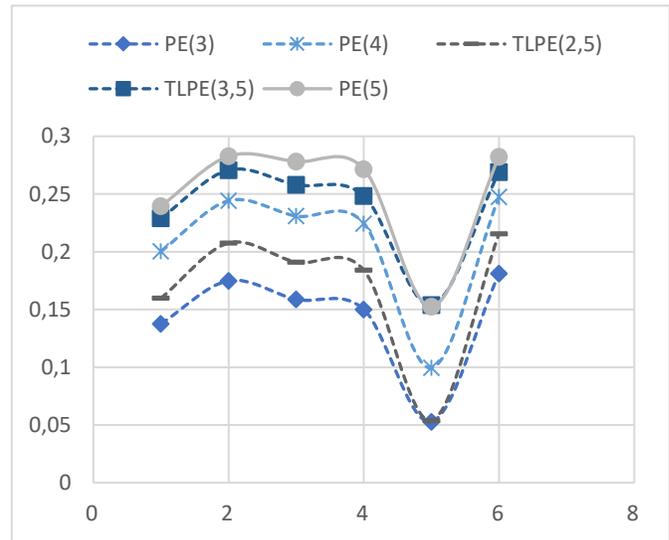

*Figure 20 - Jensen-Shannon Statistical Complexity (a)*

The statistical complexities corresponding to the modified entropies are not close to those corresponding to the original entropies due to the normalizations. The distance between the statistical complexities corresponding to the modified entropies is bigger if the number of equalities is smaller (see T5 in the figures below).

The same reasoning explains the fact that the corresponding complexities for the weighted entropies have distinct but similar plottings. Notice that the weighted entropies (WPE, WmPE, WTLPE, WmTLPE) provide similar Jensen-Shannon statistical complexity ordering. That, in our opinion, is more accurate and coherent than the values obtained with the nonweighted entropies.

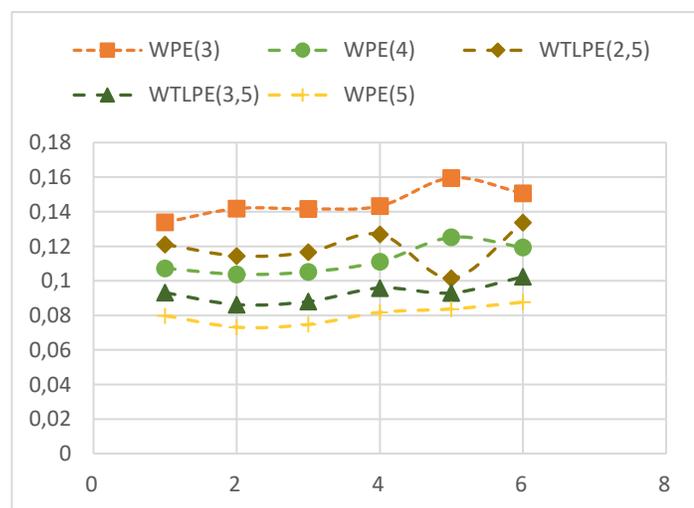

*Figure 21 - LMC Statistical Complexity (b)*

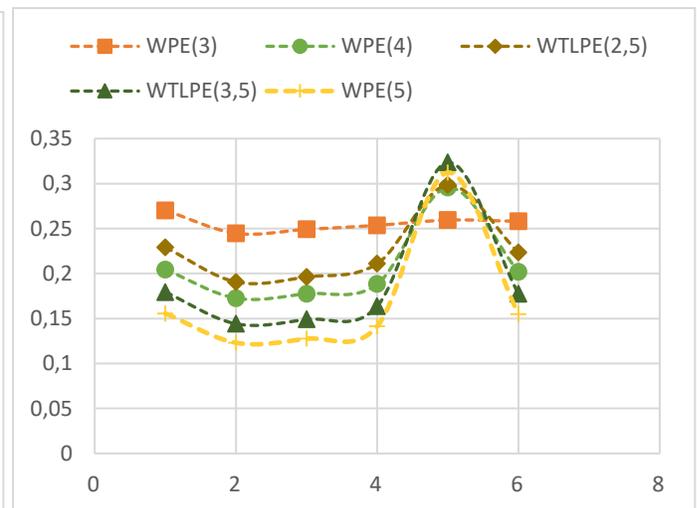

*Figure 22 - Jensen-Shannon Statistical Complexity (b)*

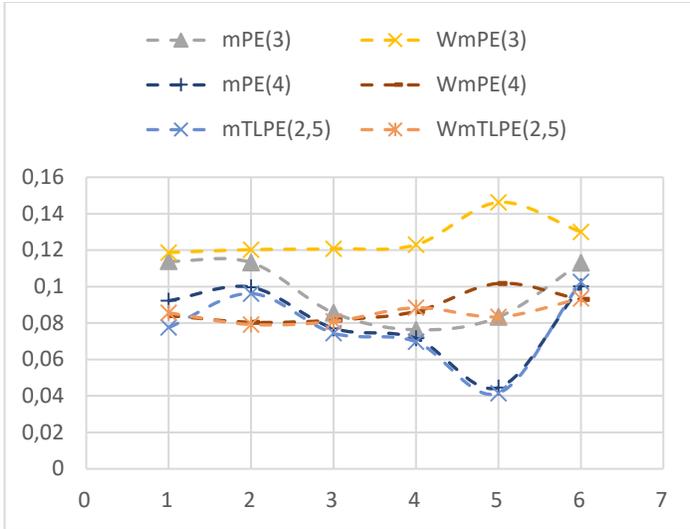
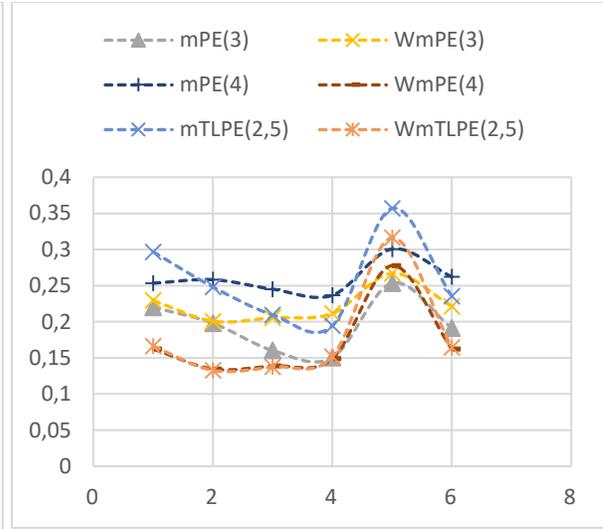

*Figure 23 - LMC Statistical Complexity (c)*     *Figure 24 - Jensen-Shannon Statistical Complexity (c)*

Shortly, a bigger number of equalities leads to closer statistical complexities for PE and mPE, respectively TLPE and mTLPE, a fact which agrees with the purpose of the mPE-algorithm, to be used when the number of equalities is big enough to make the PE-algorithm inefficient. We consider that for fire experimental data one should use the PE or mPE-algorithm (but not both), depending on the amount of equalities, if any, which occur and the embedding dimension. Further studies might attempt to establish a *threshold percentage* of tuples with equalities from the total number of tuples above which one should rather use the mPE-algorithm.

From our experimental data we see that the mPE-algorithm is not efficient here, therefore it is sufficient to avoid the equalities using the chronological ranking, without altering the conclusions.

## 4. Conclusions

From the above analysis, it can be seen that the permutation/encoding type entropies can be successfully used to detect unusual data and to perform relevant analysis of fire experiments. The results should be carefully interpreted whenever working on experimental data (subject to systematic and random errors, not to mention other unknown factors affecting the values, the blind use of the algorithms fails in some instances).

We have posed some open problems and research directions that would help researchers to choose the type of entropy to be used according to the size and other characteristics of their data (for instance the number of equalities, the time interval), improving the salient features detection.

According to our findings, as expected intuitively, the modified entropies should be used only when the amount of data is large enough and the number of equalities makes the results more relevant than those obtained with the original entropy. We conclude that they are not suited to our experimental data, which only has a small number of suitable equalities compared to the size of the data.

The new proposed entropies (WmPE, WTLPE, mTLPE, WmTLPE) are introduced here not for replacing the usual permutation/encoding type entropies, but to complement and validate the information provided by them. The weighted entropies have always provided clearer and more accurate results, which make us think that this indicates that they might be more suitable for the analysis of the data collected from fire experiments.

The results we obtained using the permutation type entropies, the statistical complexity measures as well as the weak correlation observed for T5 might indicate a turbulence (or a misfunctioning of the device); perhaps a computer simulation would help to establish a reliable explanation, however it is beyond the scope of the present paper to discuss it in detail.

***Acknowledgement*** This work was supported by a grant of the Romanian Ministry of Research and Innovation, CCCDI - UEFISCDI, project number PN-III-P1-1.2-PCCDI-2017-0350 / 38PCCDI within PNCDI III.